# A Novel Magnetoscan Setup

**S. Haindl**

*Atomic Institute of the Austrian Universities, Stadionallee 2, 1020 Vienna, Austria*

**Abstract**. Due to a modification in the original magnetoscan setup, a significant improvement in resolution was obtained. The paper focuses on experimental results which should support the idea of the new setup using two magnets with opposite direction of magnetization. This contribution to the characterization techniques of melt-grown bulk superconductors should promote the easy installation of this technique in industry. The improved magnetoscan technique may further help to investigate growth-induced inhomogeneities of the top-seeding-melt-growth process in the submillimeter range, and it offers new possibilities to the characterization of smaller structures such as superconducting films or coated conductors.

## 1. Introduction

Both, method and principles of the magnetoscan technique were introduced in [1]. It is used to detect local inhomogeneities in the shielding currents in a 1 – 2 mm layer of the surface of bulk superconductors and has certain advantages compared to conventional Hall scans of the trapped flux distribution. The setup itself represents a modification of the conventional Hall probe mapping, where the superconducting monolith is scanned by a Hall probe after magnetization. During the magnetoscan, however, a small permanent magnet scans the surface of the bulk simultaneously with the Hall probe, i. e. a magnetization of the superconductor occurs locally, restricted in horizontal as well as in vertical direction (Fig. 1a). The obtained results of the magnetoscan cover structures in the millimeter- and even in the submillimeter range, depending on their influence on local supercurrent paths. It was already shown in [2-4] that the fourfold symmetry of the growth of melt textured high temperature superconductors (i.e. the *c*-growth sector with its weak superconducting properties compared to the *a*-growth sectors including the sector boundaries), and other local weaknesses like grainboundaries and cracks can be detected. This experimental method can be used therefore in industrial applications, where a fast, cheap and non-destructive characterization is necessary, not only in view of bulk superconductors. The technique could be reproduced succesfully by other laboratories [5,6]. However, the problem of a "weak" spatial resolution remained, especially for bulk sample diameters below 20 mm and also in comparison of the results with optical micrographs in the micrometer range. Since then, efforts to improve the technique were undertaken [7]. To obtain a higher spatial resolution, changes in the setup of the original magnetoscan were tested. Firstly, the size of the magnet was reduced with the aim to receive a higher local information on the superconducting properties (Fig 1b). Secondly, a succesful modification of the magnetoscan setup can be presented: Instead of one magnet, two magnets with opposite direction of magnetization are used (Fig. 1c). Results obtained within this new setup are compared to results of the original magnetoscan, showing, in general, an accentuation of defect parts with poor superconducting properties in the surface of the bulk superconductor.

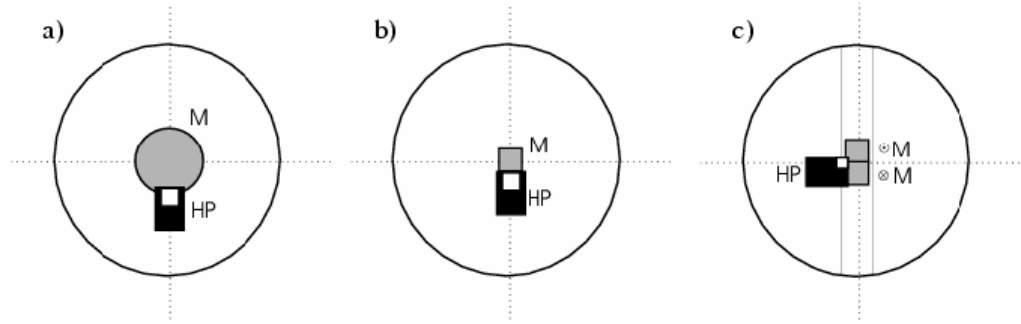

Fig. 1: Bottom view of the different Hall probe/magnet systems. a) Original magnetoscan setup, b) magnetoscan setup with smaller magnet size and c) modified magnetoscan setup with two magnets. The Hall probe is denoted by "HP", the magnet by "M".

## 2. Experimental Details

For the original magnetoscan setup a detailed description can be found in [1]. The difference in all three setups is already shown schematically in Fig. 1. For the new setup (Fig. 1c), two SmCo permanent magnets of cuboid shape were used. They were mounted side by side with opposite direction of magnetization. The Hall probe was mounted with the active area between both magnets as indicated in Fig. 1c. The new setup therefore uses the local change in direction of magnetic induction influencing the surface of the scanned monolith. The surface of the superconducting samples was scanned in a liquid nitrogen bath with a spacing of 0.2 mm between the sample surface and the Hall probe in order not to damage the active area of the probe. The scans were carried out on a computer controlled *x-y*-positioning system. Scan grids of $0.25 \times 0.25$ mm² and $0.20 \times 0.20$ mm² were used. The Hall voltage was measured by a Keithley DMM, the Hall probe current of 10 mA was supplied by a Lake Shore constant current source. Within the new setup, two possible configurations can be adjusted: *i)* both magnets parallel to the scan direction (Fig. 2a) and *ii)* both magnets serial to the scan direction (Fig. 2b). The "fingerprint" of the magnetic system on the surface of a superconducting bulk is shown in Fig. 2c. The two magnets were moved vertically down to the surface of the superconductor after cooling it (zfc) in liquid nitrogen. After a few seconds they were removed again in positive *z*-direction. Then the remaining trapped field was scanned by a HHP-SF Hall probe from AREPOC with a scan grid of $0.25 \times 0.25$ mm². Magnetic flux is trapped showing a dipole pattern inside an area of $3 \times 5$ mm² on the sample surface. It is obvious that the position of the active area of the Hall probe (Fig. 1c) coincides with a position of negligible influence of the magnetic system itself, thus small variations in the local supercurrent density (e.g. at a defect site) are amplified.

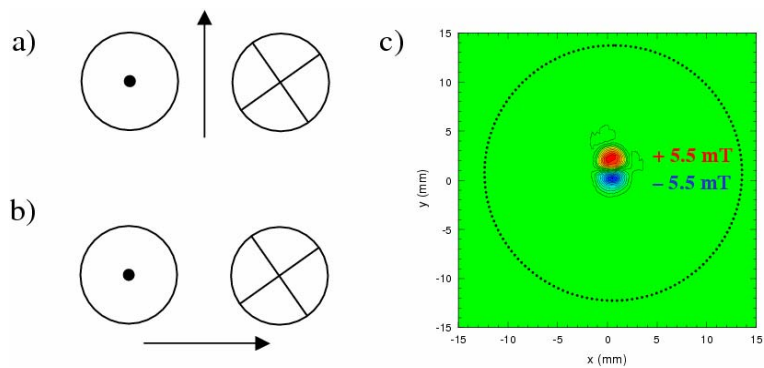

Fig. 2: Two possible configurations of the modified Magnetoscan setup. a) Magnets parallel to the scan direction and b) magnets in series to the scan direction. c) "Fingerprint" of the magnetic system on a big melt grown monolith (with a diameter of 26 mm). The dotted circle indicates the sample dimensions.

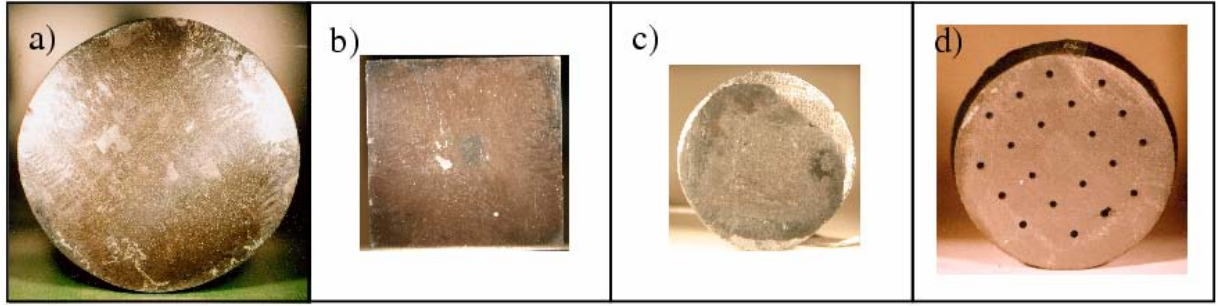

Fig. 3: Top surfaces of investigated samples: a) sample *B*, b) sample *M*, c) sample *S* and d) sample *H*.

Table 1: Technical details of both setups

|  | original setup | new setup |
|---:|:---:|:---:|
| Hall probe | AREPOC HHP-VU | ARPOC HHP-VC |
| active area | 50 × 50 μm² | 50 × 50 μm² |
| gap | 0.2 mm | 0.2 mm |
| magnet dimensions | cylindrical ⌀ = 6 mm h = 18 mm | cuboid a × b = 2 × 2 mm² c = 1 mm |
| magnet induction at the surface of the magnet | ~ 100 mT | ~ 300 mT |

For the test run of the modified magnetoscan setup top-seeded-melt-grown superconductors manufactured by different labs were investigated. A characterization of the samples can be found in Table 2. One bigger sample, *B*, one sample of medium size, *M*, one small sample, *S* and one sample with artificial holes, *H*, were tested. The top surfaces of the samples are shown in Fig. 3. Details of the top-seeding melt growth process itself can be found in literature [8-12].

Table 2: Sample characteristics

| sample code | B | M | S | H |
|---:|:---:|:---:|:---:|:---:|
| diameter | 26.4 mm | 19 × 18.5 mm² | 17.5 mm | 20.8 mm |
| thickness | 9.0 mm | 14.6 mm | 5.2 mm | 7.6 mm |
| composition | Li-doped YBCO | U-doped YBCO | Cu-Hf-doped YBCO | YBCO with drilled holes |
| additional information | cylindrical | cuboid | cylindrical | hole ⌀: 0.7 mm |
| manufacturer | IFW Dresden | IRC Cambridge | | CRISMAT-ENSICAEN |

## 3. Results and Discussion

*3.1. General comments*

The whole setup of the magnetoscan is, as already pointed out in [1], very sensitive to the parallel alignment of the Hall probe/magnet system to the surface of the superconductor. With the new setup using two magnets, the sensitivity is even larger, because of the smaller magnet size. The larger the vertical distance of the Hall probe/magnet system from the surface of the bulk, the higher the loss in resolution, i.e. the smaller the spectrum of the signal was observed. Highly impressive results were found by employing the configuration where the two magnets scan the surface of the superconductor in series. The results obtained by the different sketched magnetoscan setups shown in Fig.1 are compared in Fig. 4. The top surface of sample *M* was scanned in Fig. 4a by the original magnetoscan setup, in Fig. 4b by the setup containing the smaller magnet, in Fig. 4c by the new setup with both magnets parallel and in Fig. 4d by the new setup with both magnets in series. Comparing Fig. 4a and Fig. 4b, a clear improvement of resolution is made by using a smaller magnet. Defects in the surface of the bulk superconductor can be located with higher accuracy in Fig. 4b. The employment of two magnets in serial direction (Fig. 4d), however, improves the resolution of the magnetoscan further. Defects, such as small cracks for example, emerge stronger than in Fig. 4b. The comparison (Fig. 4c and Fig. 4d) between the two different configurations shown in Fig. 2, favours apparently the mounting of the magnets in series to the scan direction. It can also be observed, that nearly the whole sample surface is mapped in Fig. 4d), whereas in all other setups a "loss" of information occurs at the edge of the sample. Due to the higher locality achieved by using two magnets, also superconducting bulks of smaller size could be investigated with higher accuracy, which must be seen as an advance of the new setup.

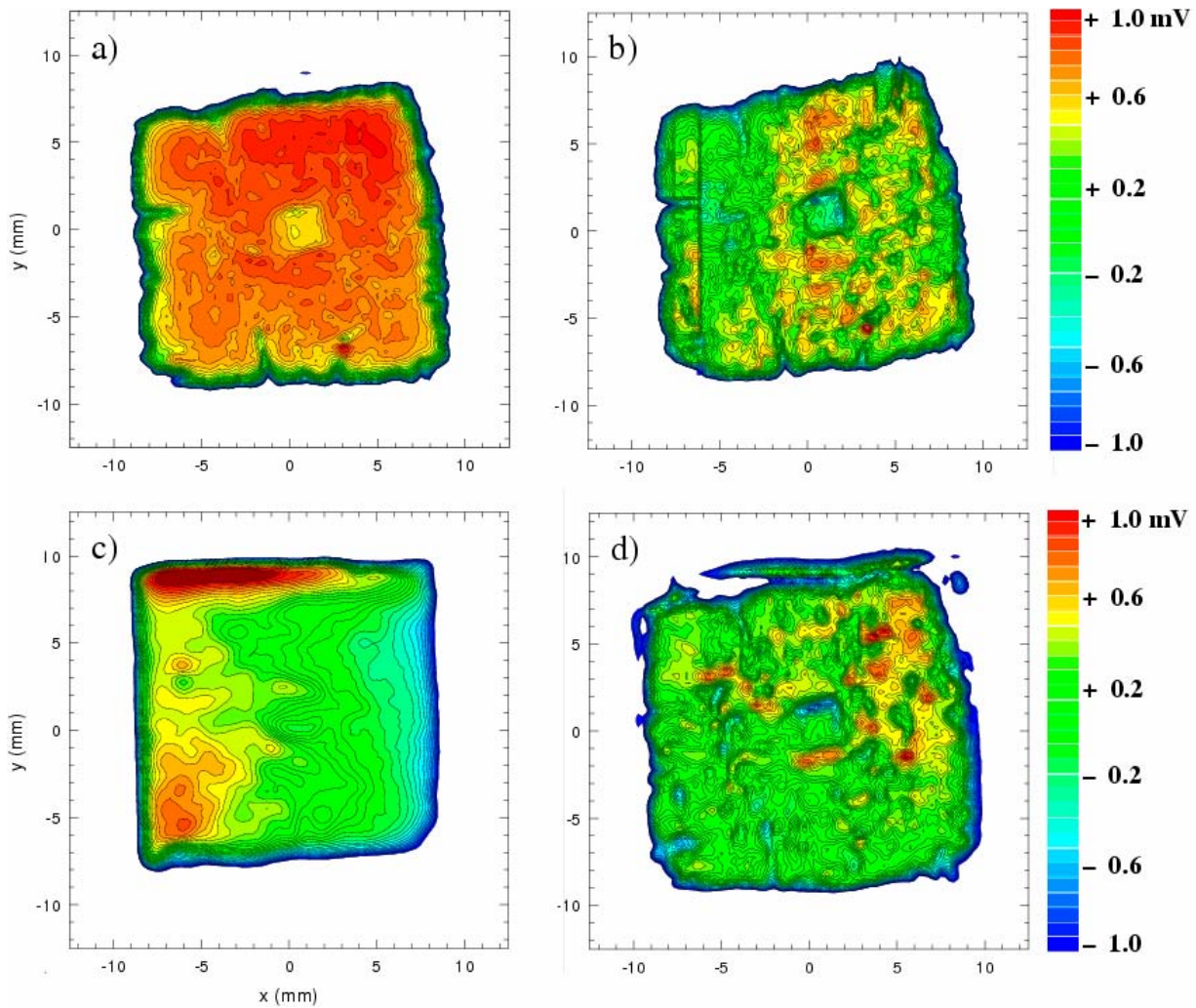

Fig. 4: Comparison between the results of different magnetoscan setups scanning sample *M*. a) Use of the original setup shown in Fig. 1a. b) Use of the setup with smaller magnet size shown in Fig. 1b. c) Use of the setup referring to Fig. 1c with two magnets parallel to the scan direction and d) use of the setup referring to Fig. 1c with two magnets serial to the scan direction.

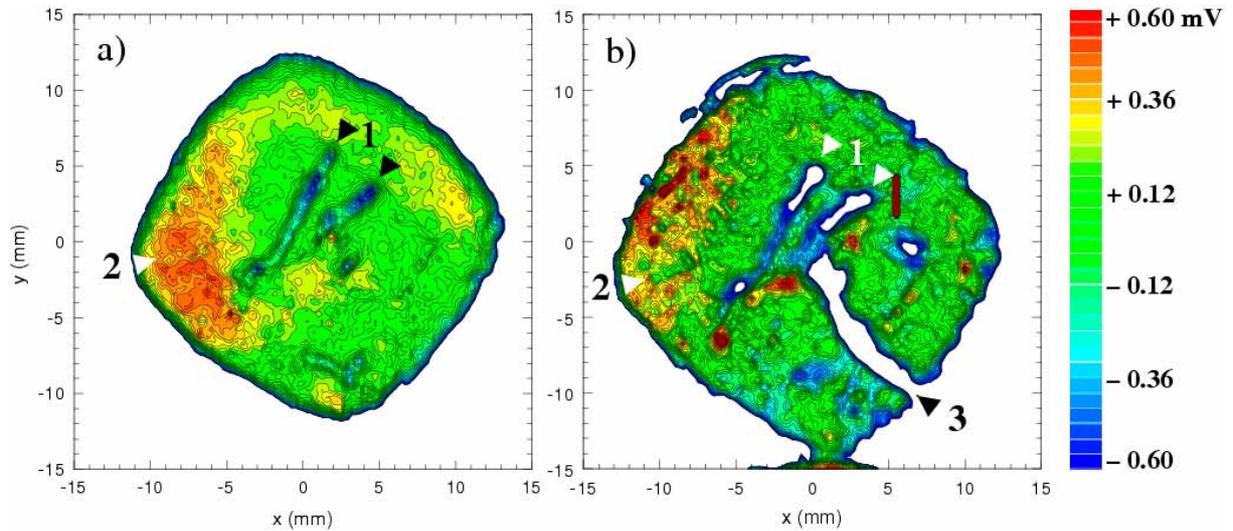

Fig. 5: Magnetoscans of sample *B*, a Li-doped YBCO bulk. Comparison between a) the original setup and b) the new setup using two magnets.

3.2. Examples of local mapping

The following examples demonstrate the potential of the improved magnetoscan. If not mentioned explicitly, the convention in the next figures will be the following: the result in a) was obtained by using the setup shown in Fig. 1a, whereas the result in b) was obtained by using the new setup introduced in Fig. 1c with the magnets in series to the scan direction. The signal spectrum for both scans were brought into congruence.

Fig. 5 represents magnetoscans of a Li-doped YBCO, sample *B*, after having removed 3.4 mm in thickness. The original sample thickness was 9.0 mm. Therefore, the *c*-growth sector appears relatively large and covers almost 50 % of the sample surface. Whereas the original magnetoscan maps the *a*-growth sectors with higher signals, the new setup maps more effectively the *a-c*-growth sector boundary. Massive cracks, which appear inside the *c*-growth sector can be resolved very well ("1"). The new setup shows a denser distribution of the contour lines within the same signal spectrum. Thus, peaks and valleys are resolved stronger ("2"). The smallest spatial signal variations in this representation are in the range of 100 μm – 250 μm. An interesting question concerns therefore the origin of these small variations, which are below 0.1 mV (which would correspond to a magnetic induction of roughly 1 mT). Because of their unlimited appearence across the whole sample surface, one suggested origin may be the *c*-macrocrack network. The *c*-macrocrack spacing was found to be around 35 μm [13]. A detailed investigation of this possibility is carried out at the moment [14].

Due to frequent cooling cycles between both measurements, a new massive crack has formed in sample *B* ("3"), shown in Fig. 5b. (The bold line-like high signal near "1" does not reflect sample properties but is more an artefact of the single measurement and therefore to neglect in the considerations.)

In Fig. 6 magnetoscans of a bulk YBCO superconductor with artificial holes are shown. With the new magnetoscan setup the holes of 0.7 mm diameter can be localized much better than before. At the lower edge of each hole, a local maximum is observed. So not to overcrowd the figure with additional lines, the holes are not drawn above the measurement, but a comparison with Fig. 3d will clarify the exact positions of the holes of sample *H*.

Fig. 7 represents the results of magnetoscans of a cyclic growth superconductor, sample *S*. The spectrum of the signal is equal for both, Fig. 7a and Fig. 7b. Again, the produced image with the new magnetoscan setup offers a higher resolution and a better distinction between qualitatively "better" and "worser" regions of the superconductor. Compared to the original setup, more details at the edges of the monolith are visible. Number "1" indicates sample regions where a gain in resolution is pretty noticable, number "2" indicates the clear evidence of four successive non-superconducting bands (lower signal) in a cyclic growth bulk superconductor. The superconducting bands (higher signal) are roughly 600 µm in thickness. This value is confirmed by scanning electron microscopical investigations, where the thickness of the superconducting regions measures several hundred micrometers.

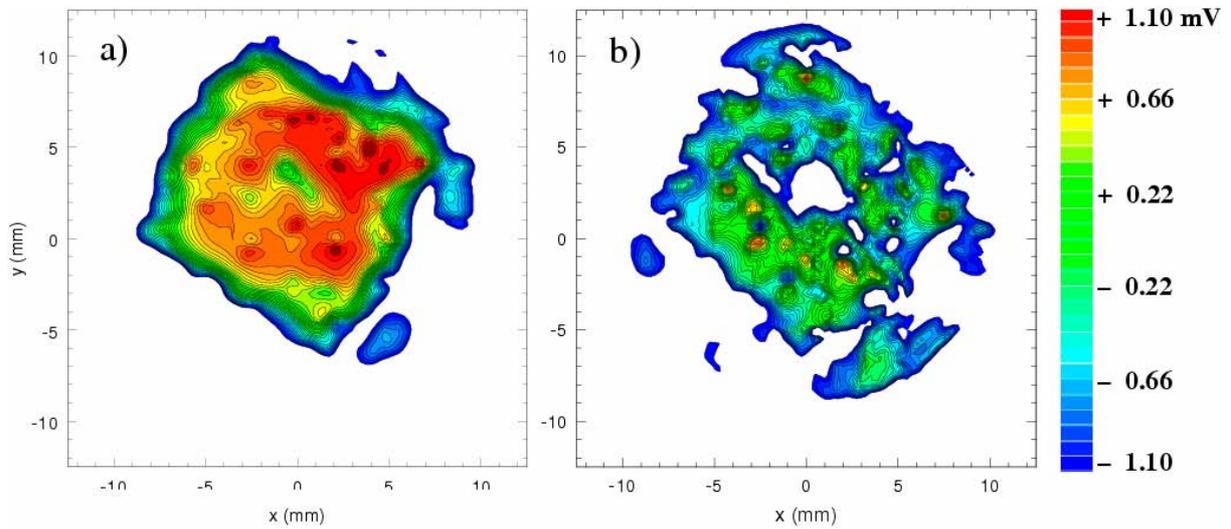

Fig.6: YBCO sample with artificial holes (sample *H*). Comparison between a) the original setup and b) the new setup using two magnets.

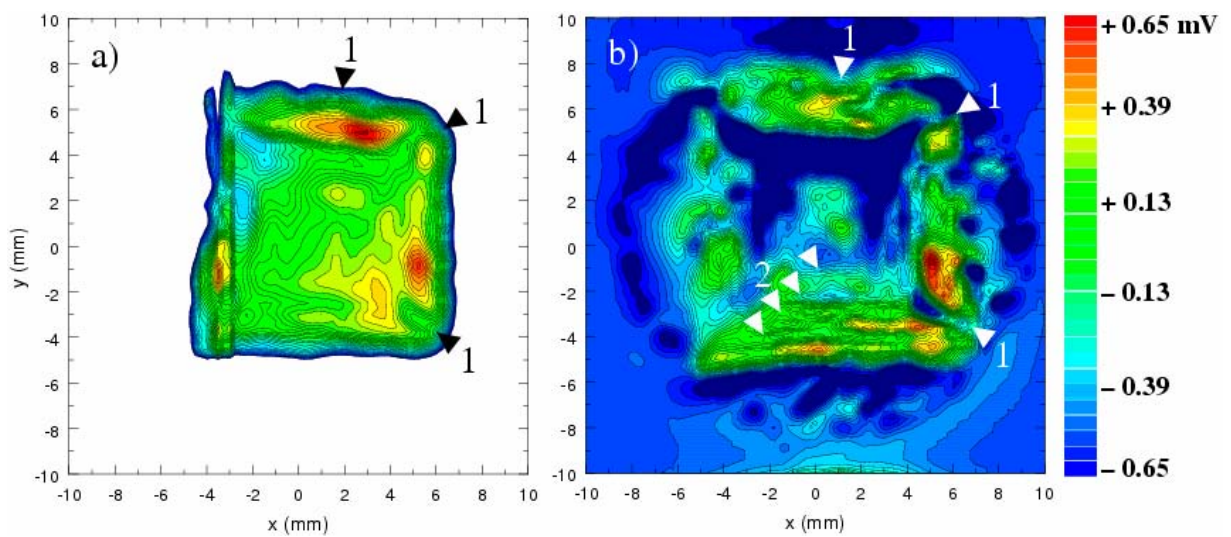

Fig. 7: Magnetoscans of sample *S* showing cyclic growth bands. a) Original setup. b) New setup using two magnets.

## 4. Conclusions

It was shown that the new setup provides a higher resolution using the serial configuration: the sample surface is mapped completely (also information from the edges is obtained); local differences of the superconducting transport properties and small defects are more accentuated. The new setup allows to resolve structures with dimensions of few 100 µm well. It is definitely more sensitive to local inhomogeneities than the original setup. Impressive results were presented, e.g. the structure of the periodic occuring bands in cyclic growth superconductors and a detailed local resolution around artificial holes in YBCO monodomains. The new setup allows to investigate also superconductors of smaller size with higher accuracy. Due to the small dimensions of the magnets, the method stays sensitive with respect to the vertical distance between the Hall probe/magnet system and the superconducting surface. More experimental work combined with analytical calculations can improve the setup further.

**Acknowledgments**
I would like to thank L. Shlyk and G. Krabbes (IFW Dresden), J. Noudem (CRISMAT-ENSICAEN Caen), N. Hari Babu and D. A. Cardwell (IRC Cambridge) for sample preparation. I am grateful to H. Hartmann and E. Tischler for technical support. This work has been supported by the Austrian Science Fund (project no. 17443).